\documentclass[letterpaper]{article} 
\usepackage{aaai2026}  
\usepackage{times}  
\usepackage{helvet}  
\usepackage{courier}  
\usepackage[hyphens]{url}  
\usepackage{graphicx} 
\usepackage{natbib}  
\usepackage{caption} 
\usepackage{booktabs} 
\usepackage{algorithm}
\usepackage{algpseudocode}

\usepackage{newfloat}
\usepackage{listings}
\DeclareCaptionStyle{ruled}{labelfont=normalfont,labelsep=colon,strut=off} 
\floatstyle{ruled}
\newfloat{listing}{tb}{lst}{}
\floatname{listing}{Listing}
\usepackage{hyperref} 
\usepackage{navigator} 
\nocopyright 

\title{MuteSwap: Visual-informed Silent Video Identity Conversion}
\author{
    Yifan Liu,
    Yu Fang,
    Zhouhan Lin
}
\affiliations{
    yifan.liu@sjtu.edu.cn, fangyu821@163.com, lin.zhouhan@gmail.com \\
    LUMIA Lab, Shanghai Jiao Tong University

}

\usepackage{bibentry}

\usepackage{multirow}
\begin{document}

\maketitle

\begin{abstract}
Conventional voice conversion modifies voice characteristics from a source speaker to a target speaker, relying on audio input from both sides. However, this process becomes infeasible when clean audio is unavailable, such as in silent videos or noisy environments. In this work, we focus on the task of Silent Face-based Voice Conversion (\textbf{SFVC}), which does voice conversion entirely from visual inputs. i.e., given images of a target speaker and a silent video of a source speaker containing lip motion, SFVC generates speech aligning the identity of the target speaker while preserving the speech content in the source silent video. As this task requires generating intelligible speech and converting identity using only visual cues, it is particularly challenging. To address this, we introduce \textbf{MuteSwap}, a novel framework that employs contrastive learning to align cross-modality identities and minimize mutual information to separate shared visual features. Experimental results show that MuteSwap achieves impressive performance in both speech synthesis and identity conversion, especially under noisy conditions where methods dependent on audio input fail to produce intelligible results, demonstrating both the effectiveness of our training approach and the feasibility of SFVC.
\end{abstract}

\begin{links}
    \link{Demo}{https://pussycat0700.github.io/MuteSwap-Demo}
    \link{Code}{https://github.com/PussyCat0700/DiVISe}
\end{links}

\section{Introduction}

Voice conversion (VC) ~\cite{wang21n_vqmivc_interspeech, freevc_iccasp23} aims to synthesize speech that preserves the linguistic content of a source utterance and adopting the vocal characteristics of a target speaker.
While traditional VC methods rely entirely on speech to extract both content reference and speaker identity, facial features have been shown to correlate closely with vocal characteristics~\cite{saeed2021FOP, hannan2025PAEFFprecisealignmentenhanced}, making them a plausible alternative for identity representation in speech conversion.
To address this, face-based voice conversion (FVC) \cite{FacebasedVC_mm21_tw,SPFaceVC_AAAI23,FVMVC,lee_HearYourVoice_F0est_interspeech24,rong_IDFaceVC_Arxiv2024} extends traditional voice conversion by extracting speaker identity from speaker images, instead of speech recordings. While FVC is an exciting advancement, it is not feasible in scenarios where source speech is noisy, corrupted, or entirely missing.

\begin{figure}[ht]
  \centering
  \includegraphics[width=\linewidth]{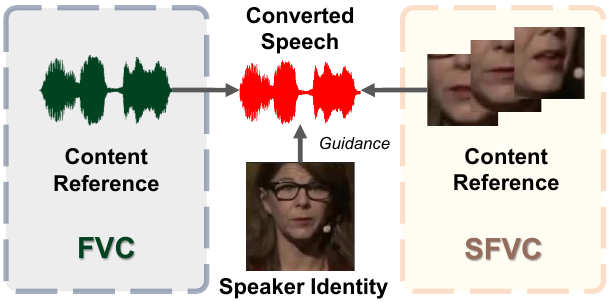}
  \caption{FVC (left) and the proposed SFVC task (right).}
  \label{fig:whatisSFVC}
\end{figure}

\begin{figure*}[t]
    \centering
    \includegraphics[width=\linewidth]{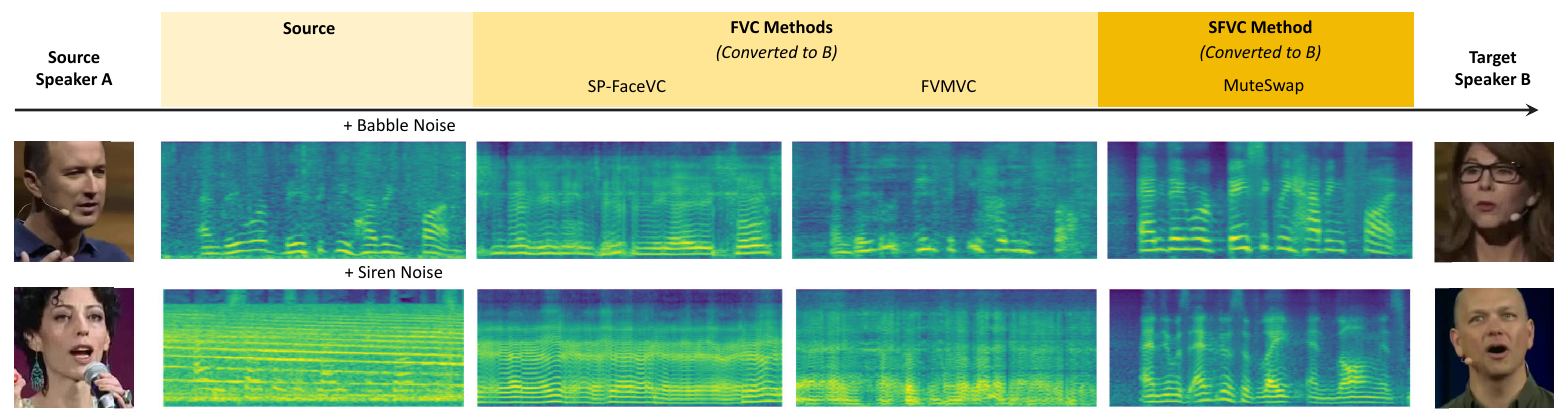}
    \caption{Comparison of log Mel-spectrograms converted from video sources with noisy sound track using FVC and SFVC. Top: source speech with babble noise. Bottom: source speech with siren noise.}
    \label{fig:fvc_vs_sfvc}
\end{figure*}

To overcome these limitations, we introduce \textbf{S}ilent \textbf{F}ace-based \textbf{V}oice \textbf{C}onversion (SFVC), a novel extension of FVC that further replaces the content reference from audio to silent lip-region video (shown in Figure~\ref{fig:whatisSFVC}). In SFVC, both content reference and speaker identity are derived from silent videos, enabling speech generation even when no audio is available. The model infers linguistic content directly from the video frames while ensuring that the synthesized speech aligns with the target speaker's identity, therefore immune to any disruption from the audio source.
We demonstrate the robustness of SFVC compared to FVC methods in Figure~\ref{fig:fvc_vs_sfvc}. As background noise intensifies from babble to siren, the performance of FVC methods declines significantly, with the converted speech exhibiting increasing spectral distortion and a progressive loss of linguistic content until it becomes unintelligible.
In contrast, SFVC consistently produces clear and intelligible speech by relying exclusively on lip movements and facial images from silent videos, making it inherently immune to audio corruption. This highlights the fundamental advantage of SFVC in achieving robust speech conversion under real-world conditions where audio is degraded or unavailable.

Compared to FVC, SFVC poses a more challenging problem: while FVC relies on high-quality speech recordings as content references, SFVC must infer linguistic content purely from silent lip-region video. This significantly increases the difficulty of content extraction, as visual cues alone provide limited and ambiguous information compared to audio. To address the challenge of content synthesis, SFVC builds on advances in video-to-speech (V2S) synthesis \cite{prajwalLip2wav, mira2022svts, Kim2023MultiTaskL2S, choi2023intelligible, revisewnhu, liu2025divisedirectvisualinputspeech}, which serves as a foundational technology for generating intelligible speech from silent videos.
However, V2S itself is not designed to handle identity conversion. Existing V2S systems typically synthesize speech in the identity of the input video’s speaker and lack the capacity to manipulate speaker characteristics independently. In this regard, our SFVC task goes beyond basic V2S by introducing speaker identity conversion as an integrated downstream objective. To the best of our knowledge, prior work has rarely explored identity conversion in the context of V2S, making SFVC one of the first to expand V2S into downstream speech manipulation tasks. This not only elevates the complexity of the task by requiring both accurate speech generation and in-place identity transformation, but also broadens the practical potential of V2S-based systems in real-world applications where content and speaker identity must be independently controlled.

A key challenge in SFVC is the lack of parallel data: real-world datasets seldom provide silent videos paired with identity-transferred speech from the same content, making fully supervised training for speaker conversion impractical. As a result, SFVC must support zero-shot identity conversion, where the model generalizes to unseen target identities without access to paired training samples. This constraint introduces two unique training challenges specific to SFVC.
The first challenge is \textbf{in-modality disentanglement} between the facial speaker identity and the lip-regional content reference. Since the video frames used for content extraction are cropped from facial image frames used for speaker identity, there is a significant overlap in visual information. This overlap makes it harder for the model to disentangle content and identity compared to FVC, where the two are provided through clearly distinct modalities.
The second challenge is the \textbf{cross-modality gap} between visual inputs and speech outputs. Since SFVC relies only on facial features to guide speaker identity in the synthesized speech, the model must learn to align visual and audio modalities. Without dedicated training to address this discrepancy, it is difficult to ensure that the facial identity effectively guides speech synthesis in a natural-sounding manner during speech conversion.

To address these challenges, we propose \textbf{MuteSwap}, the first identity conversion methodology for SFVC that converts silent facial videos into intelligible speech while simultaneously performing zero-shot speaker identity conversion based solely on reference facial images. To address the \textbf{in-modality disentanglement} challenge, a log-likelihood estimator is integrated in training with an Expectation Maximization (EM) learning process ~\cite{cheng2020clubcontrastivelogratioupper}, which minimizes the mutual information upper bound between the output embedding of facial identity encoder and lip-regional content encoder inspired by \citet{wang21n_vqmivc_interspeech}. For the \textbf{cross-modality matching} challenge, a speech encoder is trained jointly with our facial encoder using a contrastive learning approach, which ensures consistency between the speaker identity representations in the visual and audio modalities similar to CLIP ~\cite{radford2021CLIPlearningtransferablevisualmodels}. Both tasks are integrated into a single-stage training framework, allowing them to be learned simultaneously with the V2S task.

Experiments show that MuteSwap offers substantial gains over traditional FVC methods in noisy settings, maintaining high speech quality and intelligibility where audio-based methods fail. Even under clean conditions, it achieves identity conversion performance comparable to or better than FVC methods, while subjective evaluations favor MuteSwap for producing more natural and preferred speech. Furthermore, MuteSwap can function effectively as a vanilla video-to-speech system, preserving both intelligibility and visual coherence even without identity conversion.

The main contributions of this work are as follows:







\begin{enumerate} 
\item We introduce the task of Silent Face-based Voice Conversion (SFVC), which performs speech synthesis and in-place identity conversion from silent video, using only source lip-regional video frames and target speaker images without any acoustic input.
\item We propose MuteSwap, a novel single-stage framework for SFVC that disentangles visual identity from content and aligns audio-visual speaker representations, enabling zero-shot identity conversion.
\item MuteSwap shows robust performance across both noisy and clean conditions. It outperforms FVC methods under noise by preserving intelligibility and content accuracy, and achieving comparable or better results in clean settings, with its speech more favored by human perception.
\end{enumerate}

\section{Related Work}



\subsection{Face-based Voice Conversion}

The pioneering work for face-based voice conversion is FaceVC \cite{FacebasedVC_mm21_tw}, which proposes a three-stage pipeline without speaker disentanglement. SP-FaceVC \cite{SPFaceVC_AAAI23} then proposed a bottleneck-free cycle-consistent adversarial network that provides content embedding based on the low-pass filtered cepstrum.
FVMVC \cite{FVMVC} designs face-key and voice-value memory bank for embeddings from each modality and learns to minimize the discrepancy between training and inference in voice conversion. 
Later, HYFace \cite{lee_HearYourVoice_F0est_interspeech24} incorporated visual pitch estimation and improved conversion performance.
Content-identity disentanglement, however, is not addressed explicitly in training for FVC in prior works.
Recently, ID-FaceVC \cite{rong_IDFaceVC_Arxiv2024} proposed more fine-grained control of speaker identity and speech content by minimizing their mutual information, achieving further improvement in FVC performance.

\subsection{Video-To-Speech Synthesis}

The task of recovering speech from silent video, known as Video-to-Speech (V2S), was first introduced by \citet{prajwalLip2wav} and has since attracted significant attention. SVTS \cite{mira2022svts} proposed a V2S system trained on in-the-wild datasets that uses silent video frames and a reference speaker embedding as input to recover speech. MultiTask \cite{Kim2023MultiTaskL2S} aligned silent video with transcribed text from a speech recognition model to provide extra content supervision for V2S. IntelligibleL2S \cite{choi2023intelligible} requires the V2S model to predict self-supervised clustering units from a pre-trained model alongside Mel-spectrograms, which are then converted into waveform using a customized hybrid vocoder. DiffV2S \cite{choi2023diffv2s} introduced a two-stage pipeline, where the visual speaker embedding is first learned in contrast with the audio speaker embedding, followed by training a diffusion model conditioned on the visual embedding. LipVoicer \cite{yemini2024lipvoicergeneratingspeechsilent} leverages a lip reading model and an audio speech recognition model in the diffusion training process to convert speech from silent video. DiVISe \cite{liu2025divisedirectvisualinputspeech} focuses on addressing the loss of speaker characteristics in a previous V2S work \cite{revisewnhu} and proposes a two-stage training pipeline: the V2S model is first trained, followed by fine-tuning a neural vocoder \cite{kong2020hifigan} on the V2S outputs to enhance speech quality. Despite these advancements, few studies have explored the downstream applications of V2S.

\section{Methodology}

\begin{figure*}[ht]
  \centering
  \includegraphics[width=\linewidth]{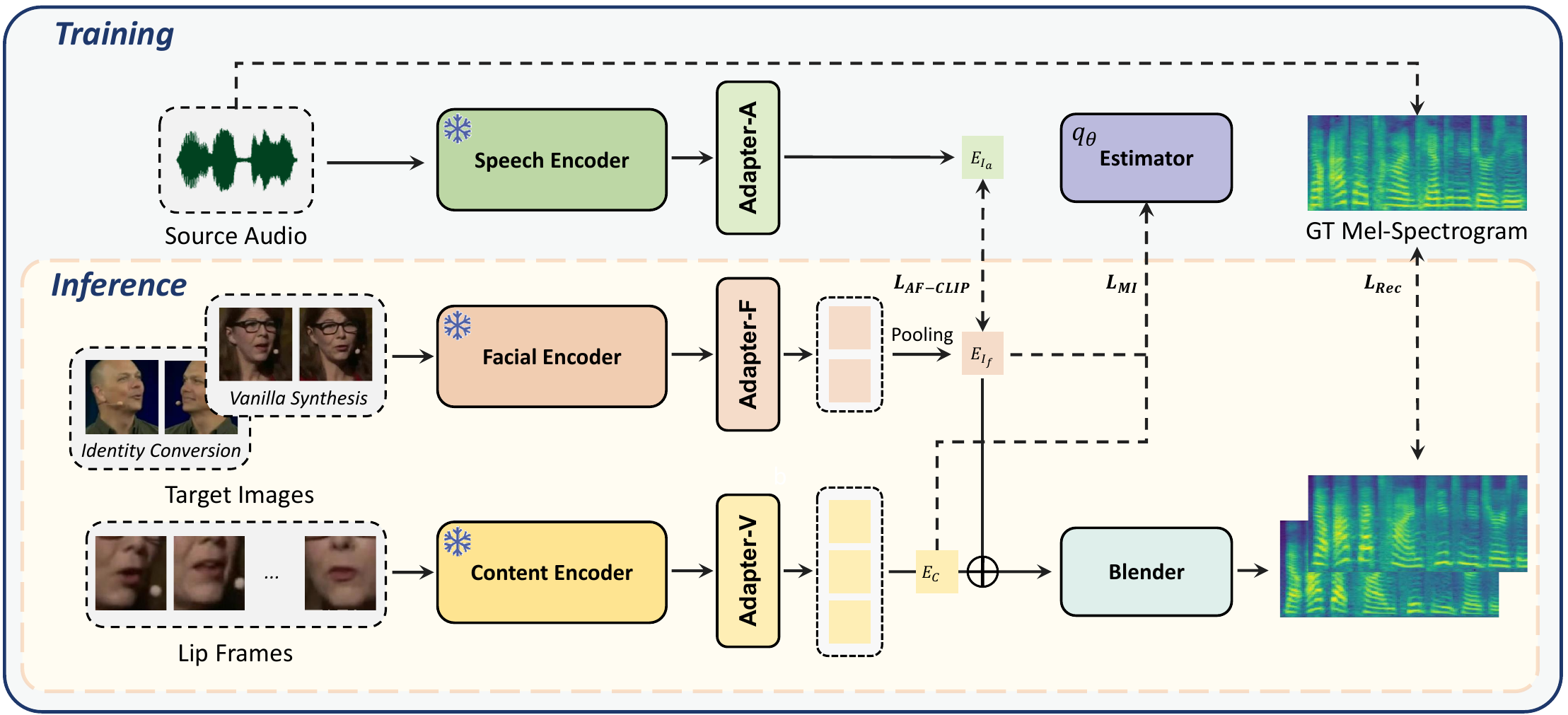}
  \caption{MuteSwap pipeline. For training, speaker identity is represented by facial images cropped from silent videos. At inference, MuteSwap is able to perform both (1) Identity conversion, when facial images are sampled from other videos; (2) Vanilla synthesis, when facial images correspond to the input video.}
  \label{fig:pipeline}
\end{figure*}

The overall architecture of MuteSwap is illustrated in Figure~\ref{fig:pipeline}. The core components for SFVC are the facial encoder and content encoder, which provide speaker identity and content references from facial images and lip-cropped video frames respectively. In training, the speech encoder is coupled with the facial encoder to handle the cross-modality gap from visual input to speech output, while the log-likelihood estimator minimizes an in-modality upper bound of mutual information between visual identity and content. To help the encoders adapt to the downstream task, we also apply trainable adapters-A/F/V before any losses are computed. Finally, the blender combines the facial identity embedding and visual content embeddings to synthesize a log Mel-spectrogram, which is fed to the fine-tuned vocoder to be converted to waveform~\cite{liu2025divisedirectvisualinputspeech}.

\begin{figure}[ht]
  \centering
  \includegraphics[width=\linewidth]{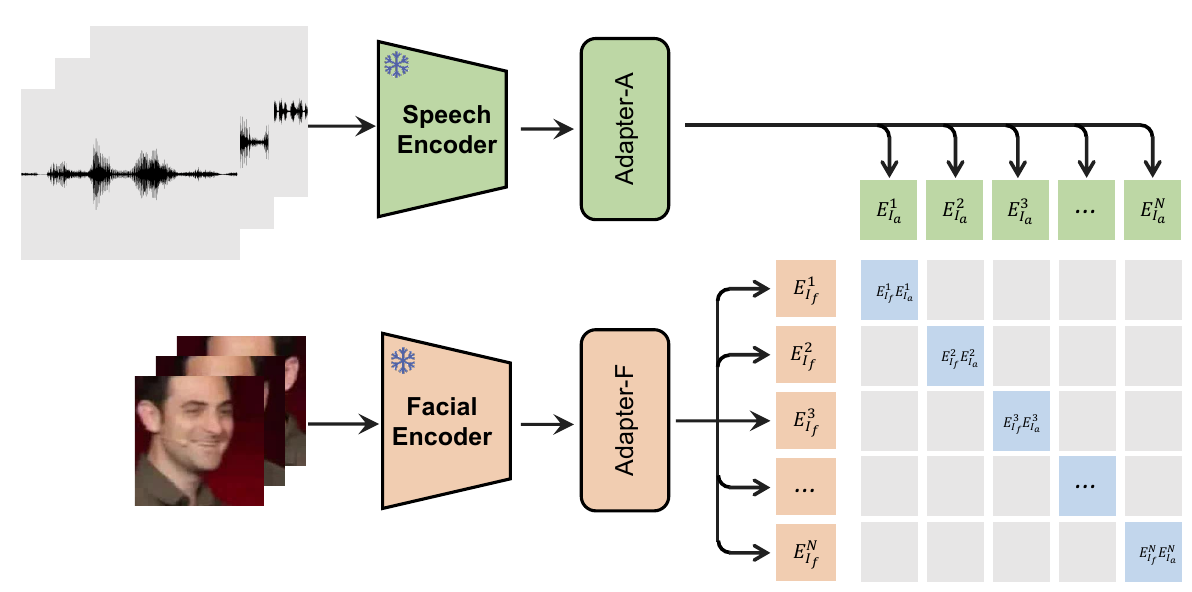}
  \caption{AF-CLIP loss}
  \label{fig:afclip}
\end{figure}

We denote the audio-based speaker identity embedding (from adapter-A) as $E_{I_a}$, the facial identity embedding (from adapter-F) as $E_{I_f}$, and the visual content embeddings (from adapter-V) as $E_C$. $E_C$ is a sequence, while $E_{I_a}$ and $E_{I_f}$ are single-vector representations per video. To obtain $E_{I_f}$, we randomly sample multiple facial images from each video, extract their embeddings via adapter-F, and apply mean pooling to aggregate them into a unified identity vector. This pooled $E_{I_f}$ is then broadcast to match the length of $E_C$ and added element-wise before being passed to the blender.

\subsection{AF-CLIP learning}

In speaker identity conversion, the facial image encoder plays a crucial role by extracting speaker-specific cues from visual appearance alone. However, visual modality alone may not fully capture how a speaker sounds, especially under identity variation. To bridge this modality gap, it is essential to align the visual speaker embedding with its audio counterpart, so that the identity cues derived from facial appearance can be more faithfully fused into the speech synthesis pipeline.

Inspired by CLIP \cite{radford2021CLIPlearningtransferablevisualmodels}, which aligns image and text representations through contrastive learning, we adopt a similar strategy to encourage alignment between facial and speech identity embeddings. Specifically, we introduce a contrastive loss, named AF-CLIP loss, that brings together facial and audio embeddings of the same speaker while pushing apart mismatched pairs within a batch.
Formally, given a batch of $N$ samples, where $E_{I_f}^i$ and $E_{I_a}^i$ denote the facial and audio identity embeddings of the $i$-th sample, the AF-CLIP loss is defined as,
\begin{equation}
    L_{A2V}=-\frac{1}{N}\sum_{i=1}^N\log\frac{\exp(\cos(E_{I_a}^i, E_{I_f}^i))}{\sum_{j=1}^N\exp(\cos(E_{I_a}^i, E_{I_f}^j))}
\end{equation}
\begin{equation}
    L_{V2A}=-\frac{1}{N}\sum_{i=1}^N\log\frac{\exp(\cos(E_{I_f}^i, E_{I_a}^i))}{\sum_{j=1}^N\exp(\cos(E_{I_f}^i, E_{I_a}^j))}
\end{equation}
\begin{equation}
    L_{AF-CLIP}=\frac{1}{2}(L_{A2V}+L_{V2A})
\end{equation}

Here, cosine similarity $\cos(\cdot, \cdot)$ serves as the metric to measure closeness between embeddings. The bi-directional loss formulation ensures that both facial and audio embeddings are jointly optimized toward a shared identity space, enforcing symmetrical alignment between the two modalities.
As illustrated in Figure \ref{fig:afclip}, the AF-CLIP loss pulls together positive pairs (samples from the same speaker, shown in blue) and pushes apart negative pairs (samples from different speakers within the same batch, shown in gray). This design promotes stronger cross-modal consistency in speaker identity encoding, which is especially critical under the SFVC setup, where speaker identity must be preserved even across modality conversion.

\subsection{Mutual Information Minimization}


A key challenge in SFVC lies in the effective disentanglement of speaker identity and spoken content, as both of them are derived from visual input. Without explicit constraints, speaker identity embedding $E_{I_f}$ and content embedding $E_C$, may share overlapping information, leading to identity leakage or content corruption in speech synthesis. To address this, we apply mutual information (MI) upper bound minimization \cite{cheng2020clubcontrastivelogratioupper}, which has been shown effective in disentangling representations in speech conversion \cite{wang21n_vqmivc_interspeech}.

Mutual information quantifies how much information is shared between two random variables. In our case, a high MI between $E_C$ and $E_{I_f}$ implies that the content embedding $E_C$ is still entangled with identity features, which compromises the effectiveness of disentangled modeling. By minimizing the upper bound of MI, we encourage $E_C$ to be independent of identity information, ensuring that all identity-relevant cues are encoded in $E_{I_f}$ for better control during speech conversion.
To achieve this, we adopt vCLUB-S \cite{cheng2020clubcontrastivelogratioupper}, which separates the optimization into two alternating steps: the E-step and the M-step, similar to the EM (Expectation-Maximization) algorithm.
In the E-step, we train a variational neural estimator $q_\theta$ to approximate the conditional distribution $p(E_{I_f}|E_C)$, by maximizing the likelihood of true sample pairs within a batch,

\begin{equation}\label{eq:l_theta} L_\theta = -\frac{1}{N}\sum_{i=1}^N\log q_\theta(E_{I_f}^i|E_C^i) \end{equation}

Intuitively, $q_\theta$ learns to assign high probability to content embeddings when conditioned on their paired identity embeddings. Once the estimator is updated, we fix its parameters and move to the M-step, where we use the learned $q_\theta$ to minimize the \textbf{MI upper bound}, as given in vCLUB-S,

\begin{equation} L_{MI} = \frac{1}{N}\sum_{i=1}^N\log\frac{q_\theta(E_{I_f}^i|E_C^i)}{q_\theta(E_{I_f}^{k_i'}|E_C^i)} \end{equation}

where $k_i'$ is uniformly sampled from $\{1,2,\dots,N\}$ to obtain a random negative index from the batch.
Inspired by \citet{wang21n_vqmivc_interspeech}, we parameterize $q_\theta(E_{I_f}|E_C)$ as a variational Gaussian estimator to approximate the conditional distribution $p(E_{I_f}|E_C)$. Since $E_C$ is a sequence while $E_{I_f}$ is a single vector, we encode $E_C$ with an LSTM and use its final hidden state as input in $q_\theta$,

\begin{equation}
    q_\theta(E_{I_f}^i|E_C^i)=\frac{1}{2}(-\frac{(E_{I_f}^i-\mu_\theta(E_C^i))^2}{\exp({\log\sigma^2_\theta(E_C^i)})}-\log \sigma_\theta^2(E_C^i))
\end{equation}

where $\mu_\theta$ and $\log\sigma^2_\theta$ are components in $q_\theta$ to estimate the Gaussian mean and log variance given $E_C$, respectively.
We alternate between these two steps during training. Over time, this encourages $E_{I_f}$ to become invariant to $E_C$, thereby achieving a stronger modular separation between content and speaker identity. This disentanglement is crucial for zero-shot SFVC, ensuring that speech is synthesized for unseen speakers without mixing content features with the speaker embedding.

\subsection{Video-to-Speech learning}

For the V2S task, we adopt the pipeline from DiVISe \cite{liu2025divisedirectvisualinputspeech}, which provides a simple and effective framework for training and evaluating V2S models. Speech reconstruction is supervised using an L1 loss between the predicted and ground-truth log Mel-spectrograms,
\begin{equation}
    L_{Rec}=\Vert\mathcal{S} - \tilde{\mathcal{S}}\Vert_1
\end{equation}
where $\mathcal{S}$ and $\tilde{\mathcal{S}}$ denote the ground-truth and synthesized log Mel-spectrograms, respectively. Given the visual content embedding $E_C$, MuteSwap generates $\tilde{\mathcal{S}}$ via the blender and optimizes this loss to learn V2S synthesis.
As noted by \citet{liu2025divisedirectvisualinputspeech}, the mismatch between $\mathcal{S}$ and $\tilde{\mathcal{S}}$ may introduce background noise when applying a vocoder pre-trained on clean audio after V2S model is trained.
To mitigate this issue, DiVISe proposes a separate vocoder fine-tuning stage, where the vocoder is further trained on V2S-synthesized spectrograms. We follow the same practice in MuteSwap by using the fine-tuned vocoder at inference time to produce cleaner speech waveforms.

\subsection{Training Strategy}

MuteSwap adopts a single-stage training framework, enabling the simultaneous optimization of content reconstruction and speaker identity disentanglement in a unified pipeline. As part of this framework, the EM update strategy used for MI minimization naturally integrates into the overall training process. Since the marginal distribution $p(E_{I_f} | E_C)$ required for MI minimization is unknown, we iteratively estimate this distribution and refine the MI upper bound in each update of the MuteSwap training pipeline. Specifically, the log-likelihood estimator $q_\theta$ first learns to approximate the conditional distribution $p(E_{I_f} | E_C)$ by minimizing the L1 loss $L_\theta$ (E-step). After $q_\theta$ is updated, it is held fixed while the rest of the MuteSwap model is optimized (M-step). During this stage, we jointly minimize the mutual information loss $L_{MI}$ together with the main task losses.

We denote the full MuteSwap model as $f_\phi$, and split its parameters into two groups: $\theta$, corresponding to the estimator $q_\theta$, and $\phi' = \phi - \theta$, representing the remaining parameters in the model. This separation ensures that the estimator $q_\theta$ is updated independently, and avoids interference between its optimization and that of the main objective. The loss function for optimizing $\phi'$ is as follows,

\begin{equation} L_{\phi'} = L_{Rec} + \lambda_{CLIP} L_{AF-CLIP} + \lambda_{MI} L_{MI} \end{equation}

where $\lambda_{MI}$ and $\lambda_{CLIP}$ are hyper parameters controlling the weight of MI loss and AF-CLIP loss, respectively. The loss function for optimizing $\theta$ is provided in Equation \ref{eq:l_theta}.

\begin{algorithm}
\caption{Training Process of MuteSwap}
\label{alg:training}
\begin{algorithmic}
    \Require Learning rate $\alpha$ and $\beta$
    \For{each training iteration}
        \State Update $q_\theta$ with $\theta \gets \theta - \alpha \nabla_\theta L_\theta$
        \State Update $f_{\phi'}$ with $\phi' \gets \phi' - \beta \nabla_{\phi'} L_{\phi'}$
    \EndFor
\end{algorithmic}
\end{algorithm}

\subsection{Inference}

\subsubsection{V2S Synthesis}

We denote the lip-region video frames as $C$ and facial images as $I_f$. At inference for standard V2S synthesis without swapping speaker identity, the speech synthesis process can be expressed as follows,
\begin{equation}
    \tilde{\mathcal{S}}=f_\phi(C,I_f)
\end{equation}
$\tilde{\mathcal{S}}$ is converted to waveform with the fine-tuned vocoder as in \citet{lipsyncexpert_lsemetric}. The parameters of log-likelihood estimator and speech encoder can be considered disposable in $\phi$ after training.

\subsubsection{Zero-shot SFVC}

Due to the proposed training schema, MuteSwap is enabled to achieve zero-shot SFVC by simply swapping the the speaker images used for inference in V2S synthesis. Given two different speakers $i$ and {j}, where we assume $C^i$ as speaker $i$'s lip region video and $I_{f}^j$ as speaker $j$'s facial images, we express the zero-shot conversion process as follows,

\begin{equation}
    \tilde{\mathcal{S}}_{ i \to j }=f_\phi(C^i,I_f^j)
\end{equation}

In this way, we allow MuteSwap to synthesize audio with both content inferred from speaker $i$ and speaker characteristics corresponding to speaker $j$. Therefore, traditional V2S synthesis can also be seen as a special case of SFVC where $i=j$.

\section{Experiments}

In this section, we begin by introducing the evaluation metrics and datasets in Section~\ref{sec:metrics} and~\ref{sec:data_and_preprocess}. We then present results demonstrating the advantages of SFVC over conventional FVC under noisy conditions in Section~\ref{sec:noisy_ic}, alongside evaluations of identity conversion in clean settings and V2S vanilla synthesis in Section~\ref{sec:clean_ic} and ~\ref{sec:v2s_syn}. Finally, we conduct ablation studies on training components in Section ~\ref{sec:ablation} and visualize the conversion process via identity embedding analysis in Section~\ref{sec:interp_vis}. Additional implementation and training details are provided in the appendix.

\subsection{Metrics}\label{sec:metrics}

Since SFVC performs both speech synthesis and identity conversion, we adopt separate metrics to evaluate each aspect.
For subjective MOS tests, the volunteers are asked to give their opinion score from 1 to 5. All MOS results are reported in a confidence interval of 95\%.

\paragraph{Speech Synthesis}
\begin{enumerate}
    \item \textit{Acoustic Quality:} We use the pretrained NISQA (v2.0) model~\cite{Mittag2021NISQAAD} to estimate subjective mean opinion scores on speech quality, denoted as \textbf{NISQA}.
    \item \textit{Content Accuracy:} We use Word Error Rate (\textbf{WER}) to evaluate content recovery accuracy, following~\cite{liu2025divisedirectvisualinputspeech}.
    \item \textit{Video-Speech Alignment:} To measure the alignment between video frames and generated speech, we employ the \textbf{LSE-C}/\textbf{LSE-D} scores \cite{lipsyncexpert_lsemetric}. This metric is commonly used in Video-to-Speech (V2S) tasks \cite{revisewnhu, choi2023diffv2s} to assess audio-visual synchronization.
    \item \textit{Speech Similarity:} We compute the cosine similarity between the ground truth and synthesized audio using a speaker encoder \cite{speakerencoder_GE2E}. This measure is referred to as Speaker Encoder Cosine Similarity (\textbf{SECS}) in DiffV2S \cite{choi2023diffv2s}, allowing us to assess how well the speaker identity is preserved in the synthesized audio.
    \item \textit{Subjective Test:} We ask 15 volunteers to assess speech quality, which we refer to as \textbf{MOS-SQ}. Volunteers are given synthesized speeches from different V2S methods and a reference speech, where they are required to give their opinion score on speech quality. The reference text is also provided for volunteers.
\end{enumerate}

\subsubsection{Identity Conversion}

\begin{enumerate}
    \item \textit{PSD:} We assess how distinct the converted voices are when using the same source speaker but different target speakers. Given a source video $i$ and target speaker images from $j$ and $k$, we form a negative pair $\tilde{\mathcal{S}}_{ i \to j }$ and $\tilde{\mathcal{S}}_{ i \to k }$. We average the cosine similarity among these pairs, which is termed the Paired Speaker Diversity (\textbf{PSD}) metric. Lower PSD values indicate more distinct speaker identities after conversion.
    \item \textit{PSH:} We form a positive pair $\tilde{\mathcal{S}}_{ i \to j_1 }$ and $\tilde{\mathcal{S}}_{ i \to j_2 }$ using different target videos $j_1$ and $j_2$ from the same speaker. We sample such pairs and define the average cosine similarity as Paired Speaker Homogeneity (\textbf{PSH}), where higher values indicate consistent identity.
    \item \textit{EER:} Following \citet{liu2025divisedirectvisualinputspeech,shi2022avemb}, we compute the Equal Error Rate (\textbf{EER}) using the cosine similarities of the positive and negative pairs. We treat the similarity scores as predictions from a binary classifier and the EER is then defined as the point where the false acceptance rate equals the false rejection rate in the Detection Error Tradeoff (DET) curve. Compared to average similarity metrics like PSH and PSD, EER captures the full distribution of scores and provides a more robust evaluation of speaker identity preservation and separation.
    \item \textit{Subjective Test:} We ask 15 volunteers to give face-voice matching mean opinion score (\textbf{MOS-FVM}). Given the reference facial image and utterance of the target speaker, the volunteers are required to give matching scores between the converted speech and their vocal impression of target speaker. As speech content is not relevant in this test, no reference text is provided, so that volunteers can only focus on the speaker characteristics in the test.
\end{enumerate}

\subsection{Data and Preprocessing}~\label{sec:data_and_preprocess}

We use the LRS3-TED dataset (LRS3) \cite{Afouras2018LRS3TEDAL} for both video-to-speech (V2S) synthesis and SFVC training. LRS3 contains approximately 300k video samples (25 fps) from a wide range of speakers. For training, we adopt the \textbf{LRS3 top 200} subset, which includes 25k videos from the 200 speakers with the most available data, as defined by \citet{FVMVC} to address the uneven speaker distribution in the entire dataset. For in-domain evaluation, we use the validation split from \citet{Shi2022AVHuBERT} and the original LRS3 test set. All noisy versions of the source speech used in our experiments are created by sampling noise segments from the MUSAN dataset~\cite{snyder2015musan} and mixing them with the clean source speech.
In addition to LRS3, we also assess out-of-domain generalization using the test set of the VoxCeleb2 dataset~\cite{VoxCeleb2Chung_2018}, where no fine-tuning is performed to ensure a fair cross-domain evaluation.
For positive and negative pairs in LRS3, We sample 1600 pairs each from 4 source and 8 target speakers given in \citet{FVMVC}. For pairs in VoxCeleb2, we randomly sample 6000 pairs each from its test set.

For video data preprocessing, we follow the pipeline in \cite{liu2025divisedirectvisualinputspeech} by exporting lip-regional video with 96x96 pixel regions in each frame. In training, a random crop of 88x88 is applied with a 50\% chance of horizontal flip to fit the input to content encoder. In evaluation, we simply center-crop the video.
For image data, we extract images from the original videos every 10 frames, where the number of frames is constrained to a minimum of 4 and a maximum of 40 during preprocessing. During training and evaluation, we randomly sample up to 16 images to feed into the facial encoder. When a video has fewer than 16 exported images (e.g., only 5 images), we first use all the available frames (3 images × 5 = 15) and then randomly sample the remaining images to complete the 16-frame requirement (in this case, 1 more image).

\subsection{Noisy Identity Conversion}~\label{sec:noisy_ic}

\begin{table}[ht]
\centering
\small
\begin{tabular}{ccr}
\toprule
\textbf{Method}&\textbf{NISQA}$\uparrow$&\textbf{WER(\%)}$\downarrow$\\
\hline
Source & 3.660 & 3.64 \\
Source (noisy) & 1.815 & 34.36 \\
\hline
\multicolumn{3}{c}{\textit{FVC (clean source)}} \\
SP-FaceVC & 3.771 & 74.98 \\
FVMVC & \textbf{3.620} & \textbf{27.00} \\
\hline
\multicolumn{3}{c}{\textit{FVC (noisy source)}} \\
SP-FaceVC & 2.595 & 95.47 \\
FVMVC & 2.490 & 81.55 \\
\hline
\multicolumn{3}{c}{\textit{SFVC}} \\
MuteSwap & \textbf{3.220} & \textbf{44.13} \\
\bottomrule
\end{tabular}%
\caption{Intelligibility evaluation on speech converted by SFVC and FVC.}
\label{tab:SFVC_vs_fvc_noise}
\end{table}

We evaluate the intelligibility of speech converted by FVC methods and MuteSwap in noisy conditions to demonstrate the robustness of SFVC when the source speech is degraded. As shown in Table~\ref{tab:SFVC_vs_fvc_noise}, we report only NISQA and WER, since identity conversion metrics are no longer meaningful when the generated speech is severely degraded and lacks intelligible content.

In clean settings, FVC methods perform reasonably well, with FVMVC~\cite{FVMVC} achieving a WER of 27.00\%. However, when noise is added to the source speech, their performance deteriorates sharply. Both NISQA and WER exhibit significant degradation. FVMVC's WER increases to 81.55\%, reflecting severe loss of content and intelligibility. Audio samples illustrate this sharp contrast, revealing how noise severely disrupts the generation process.

In contrast, MuteSwap remains unaffected, as it does not rely on audio input. Its performance remains stable across clean and noisy conditions, consistently achieving a NISQA of 3.220 and a WER of 44.13\%. This resilience highlights a key advantage of SFVC: by relying solely on visual input, it is immune to audio corruption and remains reliable even when speech-based methods fail.

\subsection{Clean Identity Conversion}~\label{sec:clean_ic}

\begin{table*}[ht]
\centering
\small
\begin{tabular}{ccccc|ccc}
\toprule
\multicolumn{1}{}{} & \multicolumn{4}{c}{\textit{LRS3}} & \multicolumn{3}{c}{\textit{VoxCeleb2}} \\
\textbf{Method}&\textbf{PSH}$\uparrow$&\textbf{PSD}$\downarrow$&\textbf{EER}(\%)$\downarrow$&\textbf{MOS-FVM}$\uparrow$&\textbf{PSH}$\uparrow$&\textbf{PSD}$\downarrow$&\textbf{EER}(\%)$\downarrow$\\
\hline
Reference & - & - & - & 4.85±0.06 & - & - & - \\
\hline
 &&&\textit{FVC (clean input)}&&&& \\
SP-FaceVC & 0.7763 & 0.7399 & 42.00 & 1.83±0.15 & 0.8206 & 0.7892 & 42.59 \\
FVMVC & 0.8580 & \textbf{0.6875} & 22.37 & 2.87±0.19 & 0.8349 & \textbf{0.7537} & 33.71 \\
\hline
 &&&\textit{SFVC}&&&& \\
\textbf{Baseline} & \textbf{0.9576} & 0.7900 & \underline{6.00} & 2.54±0.20 & \textbf{0.9124} & 0.8105 & \textbf{22.12} \\
MuteSwap & \underline{0.9448} & \underline{0.7292} & \textbf{4.63} & \textbf{3.09±0.20} & \underline{0.8811} & \underline{0.7588} & \underline{22.78} \\
\bottomrule
\end{tabular}%
\caption{Objective speech conversion performance comparison for SFVC and FVC. All models are trained in LRS3. Evaluation is conducted on both test sets of LRS3 (in-domain) and VoxCeleb2 (out-of-domain). "Reference" refers to the target utterance.}
\label{tab:SFVC_vs_fvc}
\end{table*}


Table~\ref{tab:SFVC_vs_fvc} compares the speech conversion performance of FVC and SFVC models on both in-domain (LRS3) and out-of-domain (VoxCeleb2) test sets, where all models are trained in LRS3. The \textbf{Baseline} for SFVC is MuteSwap trained solely with the V2S reconstruction loss, without speech encoder or log-likelihood estimator.

Although FVC models take clean speech as input and are expected to serve as an upper bound for SFVC, MuteSwap consistently achieves comparable or even better results in both datasets. In terms of objective metrics, MuteSwap achieves lower EER than all FVC methods on both LRS3 and VoxCeleb2, indicating stronger conversion performance in speaker verification. Its PSH also surpasses FVC on both datasets, indicating tighter within‑speaker consistency.
For PSD, MuteSwap falls behind FVMVC~\cite{FVMVC} on LRS3, but achieves comparable performance on VoxCeleb2, suggesting that its ability to preserve speaker diversity improves when generalizing to unseen domains.
Compared with the SFVC baseline, MuteSwap trades off a small drop in PSH for a clear reduction in PSD, while maintaining a similar EER. This indicates a better capacity to distinguish between negative pairs, which contributes to improved speaker diversity without compromising verification performance.

On subjective evaluation, MuteSwap also achieves higher MOS-FVM scores than FVC methods, showing that human listeners find its converted speech more aligned with the target identity. This further confirms the effectiveness of SFVC, despite lacking audio input. 
These results highlight the potential advantage of SFVC: since both content and identity references come from the same visual modality, the inputs are more naturally aligned. This modality harmony may lead to more coherent and perceptually convincing conversions, helping SFVC match or even outperform audio-based methods in voice conversion, in both seen and unseen domains.

\subsection{V2S Vanilla Synthesis}~\label{sec:v2s_syn}

\begin{table*}[ht]
\centering
\small
\begin{tabular}{ccc|ccc}
\toprule
\multirow{2}{*}{\textbf{Method}} & \multicolumn{2}{c}{Intelligibility} & \multicolumn{3}{c}{Consistency} \\
&\textbf{NISQA}$\uparrow$&\textbf{WER(\%)}$\downarrow$&\textbf{SECS}$\uparrow$&\textbf{LSE-C}$\uparrow$&\textbf{LSE-D}$\downarrow$\\
\hline
Reference & 3.275 & 5.58 & 1.000 & 7.627 & 6.886 \\
\hline
SVTS & 1.390 & 67.56 & 0.521 & 6.739 & 7.577 \\
MultiTask & 1.411 & 68.56 & 0.495 & 5.117 & 8.888 \\
DiffV2S & \textbf{3.483} & 46.82 & 0.609 & 7.167 & 7.271 \\
DiVISe & 3.121 & \underline{39.03} & \underline{0.629} & \underline{7.994} & \underline{6.512} \\
MuteSwap & \underline{3.133} & \textbf{38.53} & \textbf{0.630} & \textbf{8.008} & \textbf{6.472} \\
\bottomrule
\end{tabular}%
\caption{Objective intelligibility and speaker matching metrics in V2S synthesis.}
\label{tab:v2s_syn}
\end{table*}

\begin{table}[ht]
\centering
\small
\begin{tabular}{cc}
\toprule
\textbf{Method}&\textbf{MOS-SQ}\\
\hline
Reference & 4.53±0.11 \\
\hline
IntelligibleL2S & 2.53±0.18 \\
LipVoicer & 1.77±0.15 \\
DiVISe & 3.65±0.17 \\
MuteSwap & \textbf{3.85±0.14} \\
\bottomrule
\end{tabular}%
\caption{Subjective Speech Quality assessment (MOS-SQ) for V2S synthesis.}
\label{tab:mos_quality}
\end{table}

Table \ref{tab:v2s_syn} presents the performance comparison of V2S methods on the LRS3 test set. We train DiVISe \cite{liu2025divisedirectvisualinputspeech} on the same LRS3 top 200 dataset as MuteSwap, using identical training configurations but without facial image input. This ensures that the only difference between the two models lies in the presence or absence of facial guidance. For other V2S baselines, we use the publicly available audio samples provided by the authors, which are trained on significantly larger datasets, such as the entire dataset of LRS3 \cite{Kim2023MultiTaskL2S, choi2023intelligible, choi2023diffv2s} or LRS3 combined with VoxCeleb2 \cite{mira2022svts}, both of which contain 10 times more samples than LRS3 top 200 in our training.

MuteSwap achieves the best performance on WER, SECS, LSE-C, and LSE-D, indicating strong accuracy in content reconstruction, consistent speaker identity, and reliable lip-speech synchronization. It also performs competitively on NISQA, ranking just behind DiffV2S~\cite{choi2023diffv2s}. Compared with DiVISe, MuteSwap consistently outperforms it across all metrics, which indicates the improvement of V2S capability with facial guidance.
In addition, Table \ref{tab:mos_quality} reports the MOS-SQ scores on the LRS3 test set. Notably, MuteSwap ranks first among all methods, achieving the highest human-rated quality score, which suggests that the integration of facial image input along with the joint optimization for speaker identity conversion effectively improves the perceived naturalness and intelligibility of synthesized speech from silent video.

\subsection{Ablation Studies}\label{sec:ablation}

\begin{table}[ht]
\centering
\small
\begin{tabular}{lccc}
\toprule
\textbf{Model}&\textbf{PSH}$\uparrow$&\textbf{PSD}$\downarrow$&\textbf{EER(\%)}$\downarrow$\\
\hline
 MuteSwap & 0.9448 & \textbf{0.7292} & \textbf{4.63} \\
 \phantom{  } - MI Loss & 0.9574 & 0.7733 & 4.81 \\
 \phantom{Fa  } - AF-CLIP Loss & \textbf{0.9576} & 0.7900 & 6.00 \\
\bottomrule
\end{tabular}%
\caption{Ablation studies on training and components.}
\label{tab:loss_ablation}
\end{table}

Table~\ref{tab:loss_ablation} presents the results of ablation experiments on conversion metrics. We compare MuteSwap against two ablated variants: (1) the model without MI Loss (also disabling the log-likelihood estimator $q_\theta$ and the EM update used in training), and (2) the model without both MI Loss and AF-CLIP Loss (removing the speech encoder from training).

Removing the MI loss leads to a noticeable increase in PSD and a slight rise in EER, indicating degraded identity conversion quality and weaker speaker discriminability. This confirms that disentangling identity and content via MI minimization plays a central role in preserving speaker distinctiveness. When AF-CLIP loss is also removed, the degradation becomes more pronounced, especially in EER (rising from 4.81\% to 6.00\%), showing that audio-visual alignment is also essential for robust identity conversion.
Interestingly, PSH improves as these modules are removed. However, this gain could be misleading, as higher PSH in these cases likely stems from the model generating overly similar speech across different targets, at the cost of losing identity distinction. This is supported by the accompanying increase in PSD and EER.

Overall, these results emphasize that identity-content disentanglement is crucial to identity conversion performance, while AF-CLIP Loss ensures the alignment of identity across audio-visual modalities.

\subsection{Identity Embedding Interpolation}~\label{sec:interp_vis}

\begin{figure*}[ht]
  \centering
  \includegraphics[width=\linewidth]{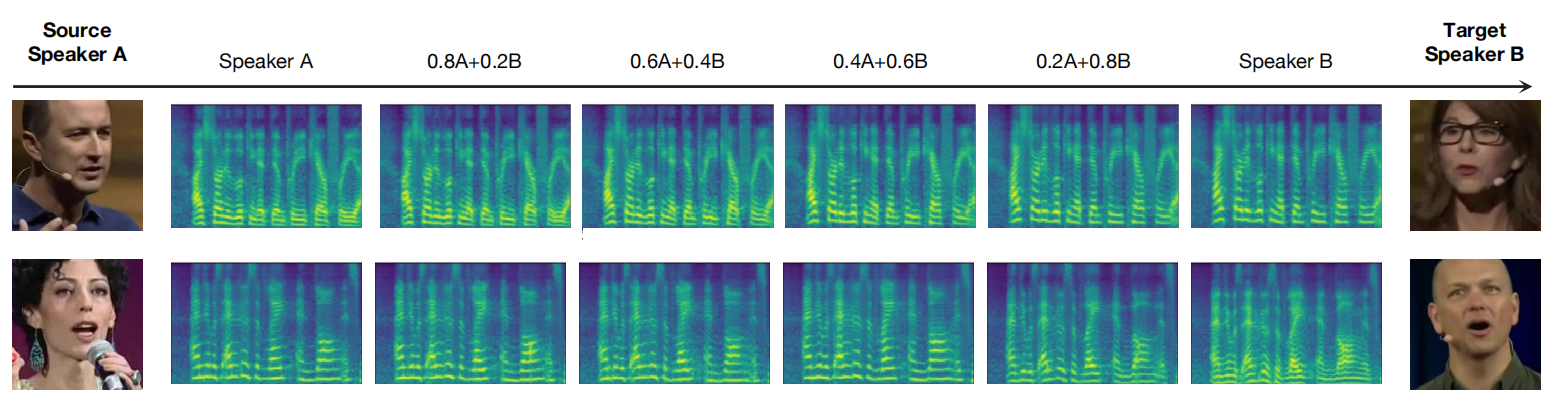} 
  \caption{Identity Embedding interpolation. The mixing ratio for target speaker facial embedding increases from left to right. }
  \label{fig:id_interp}
\end{figure*}

To further examine MuteSwap’s ability to control speaker identity in speech synthesis, we conduct an interpolation experiment on facial identity embeddings. This setup visualizes how identity conversion is realized within MuteSwap in a continuous and interpretable manner.
Given a source speaker $i$ and a target speaker $j$, we linearly interpolate their identity embeddings and synthesize speech with the mixed identity:

\begin{equation}
    \tilde{\mathcal{S}}_{ i \to \alpha j+(1-\alpha)i }=f_\phi(C^i,(1-\alpha)I_f^i+\alpha I_f^j), \alpha \in [0,1]
\end{equation}

Here, $\alpha$ controls the contribution of each speaker’s identity. When $\alpha=0$, the speech reflects the original identity of the source speaker, while $\alpha=1$ represents complete conversion to the target speaker. Intermediate values of $\alpha$ yield a gradual blend between the two.

As shown in Figure~\ref{fig:id_interp}, the synthesized log Mel-spectrograms evolve smoothly with increasing $\alpha$. Male-to-female conversions exhibit rising energy in high-frequency bands, whereas female-to-male conversions show increased concentration in lower frequencies. This smooth progression not only verifies that MuteSwap achieves disentanglement between content and identity, but also highlights its fine-grained control over speaker characteristics.

Such interpolation-based control also hints at potential applications beyond binary speaker conversion. For example, generating virtual voices with hybrid attributes or adjusting speaker identity continuously to match expressive requirements. It demonstrates that MuteSwap does not merely switch speaker identities, but embeds identity as a manipulable variable within the synthesis process.

\section{Conclusion}

In this work, we introduce the task of SFVC, which performs speech recovery and in-place speaker identity conversion using only visual input. To tackle its challenges, we propose MuteSwap, the first approach to enable speaker-content disentanglement and audio-visual identity alignment for zero-shot identity conversion without relying on acoustic input. SFVC proves especially valuable when source speech is unavailable or corrupted, as MuteSwap remains robust under noisy conditions where audio-based methods degrade significantly. Even in clean settings, MuteSwap achieves conversion quality comparable to or better than audio-driven approaches. Beyond conversion, MuteSwap also extends the capabilities of video-to-speech synthesis, laying the groundwork for future research and applications. We hope this work serves as a solid benchmark for SFVC and encourages further exploration in this direction.

\bibliography{aaai2026}

\section{Appendix}

\subsection{Modules and Training}

\subsubsection{Modules}
To accelerate convergence, we leverage several pre-trained encoders. The facial encoder is a FaRL \cite{FaRL} feature extractor\footnote{We use FaRL encoder pre-trained for 16 epochs on LAIONFace20M \cite{FaRL}.}. The visual encoder is a ResNet-18 \cite{resnet18} backbone from AV-HuBERT \cite{Shi2022AVHuBERT}\footnote{We use AV-HuBERT Large pre-trained on LRS3 \cite{Afouras2018LRS3TEDAL} and VoxCeleb2 \cite{VoxCeleb2Chung_2018}.}. The speech encoder is directly adopted from Resemblyzer \cite{speakerencoder_GE2E}. All three encoders are frozen during training.
Training updates are only applied to the subsequent modules, starting from adapter-V/A/F. For identity adaptation, both adapter-A and adapter-F are initialized as randomly projected linear layers. For content adaptation, which is more challenging due to its semantic complexity \cite{liu2025divisedirectvisualinputspeech, choi2023intelligible}, we use the Transformer encoder from AV-HuBERT, which is initialized with pre-trained parameters.
The blender module is implemented as a 4-layer Conformer \cite{Gulati2020ConformerCT} with 4 attention heads and a hidden dimension of 1024. The log-likelihood estimator uses an LSTM to summarize the content sequence via its last hidden state, which is then passed to MLPs for distribution estimation. We adopt the content encoder, adapter-V, and blender structures from DiVISe~\cite{liu2025divisedirectvisualinputspeech} for MuteSwap, preserving the core V2S capability while extending the model to enable identity conversion.

\subsubsection{Training}
We train MuteSwap on LRS3 top 200 dataset over 15k updates using a single RTX ADA6000 GPU with a batch size of 64. We use two AdamW \cite{loshchilov2017decoupledAdamW} optimizers to update the parameters of the log-likelihood estimator $q_\theta$ and other modules of MuteSwap $f_{\phi'}$, with $\beta_1=0.9$, $\beta_2=0.98$.
For $q_\theta$, we apply a constant learning rate of 1e-3. For $f_{\phi'}$, we adopt a tri-stage learning rate schedule inspired by \cite{liu2025divisedirectvisualinputspeech, revisewnhu}: the learning rate linearly increases to 1e-4 in the first 5\% of training, remains constant for the next 10\%, and then linearly decays to 5e-6 over the remaining 85\%.
For $q_\theta$, we apply a constant learning rate of 1e-3. We set $\lambda_{CLIP}=0.1$ to balance it with the reconstruction loss. We set $\lambda_{MI}=0.01$, following \cite{wang21n_vqmivc_interspeech}, as increasing this value leads to training instability.

For waveform synthesis, we use a fine-tuned vocoder to address the training-evaluation discrepancy in synthesized log Mel-spectrograms following~\cite{liu2025divisedirectvisualinputspeech}. Specifically, we fine-tune HiFi-GAN vocoder \cite{kong2020hifigan} on the output of a modified DiVISe model that incorporates our facial encoder and adapter for better acoustic quality, using 8 RTX 4090 GPUs for 45k updates with a batch size of 8 per GPU. The modified DiVISe model is trained on the LRS3~\cite{Afouras2018LRS3TEDAL} dataset for a maximum of 100k updates, using a per-device batch size of 32 on 2 RTX ADA6000 GPUs. We employ a warmup strategy where the learning rate increases linearly to 1e-4 during the first 5\% of training, followed by decay according to a scheduler, where the learning rate is reduced by a factor of 10 when no improvement in NISQA~\cite{Mittag2021NISQAAD} is observed. The evaluation is done for every 2.5k updates. The training process includes early stopping with a patience of 3, where the content encoder and adapter-V are frozen for the first 5k updates. Due to early stopping, the V2S model selected at 90k updates is used to fine-tune the vocoder.

\subsection{Additional Studies}

\subsubsection{Number of speaker images}

\begin{figure}[ht]
  \centering
  \includegraphics[width=\linewidth]{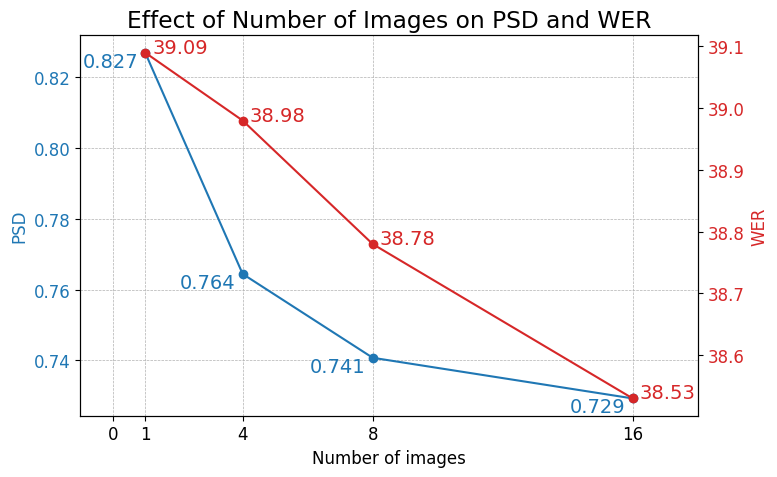}
  \caption{The effect of the number of maximum frames used for facial encoding in each video.}
  \label{fig:sup_nframes}
\end{figure}

Figure~\ref{fig:sup_nframes} presents the effect of the number of images used during training and evaluation on the performance of the MuteSwap model. We investigate how pooling image embeddings from multiple frames impacts the conversion of speaker identity, as opposed to randomly selecting just one frame. The results show that using a pool of images significantly improves the accuracy of speaker identity conversion, as reflected by the PSD metric. For instance, pooling embeddings from 4 randomly sampled images leads to a notable improvement (0.764) compared to using only a single frame (0.827).

As the number of sampled images increases, both PSD and WER metrics show consistent improvement, suggesting that more image data contributes to better speaker identity representation and content recovery. However, the rate of improvement diminishes as the number of images grows from 4 to 16, with PSD improving by only 0.035 compared to a larger improvement of 0.063 when increasing from 1 to 4 images. Given that further increases in the number of images did not yield significant gains, we chose 16 images used for both training and evaluation.

\subsubsection{DET Curves}
\begin{figure}[ht]
  \centering
  \includegraphics[width=\linewidth]{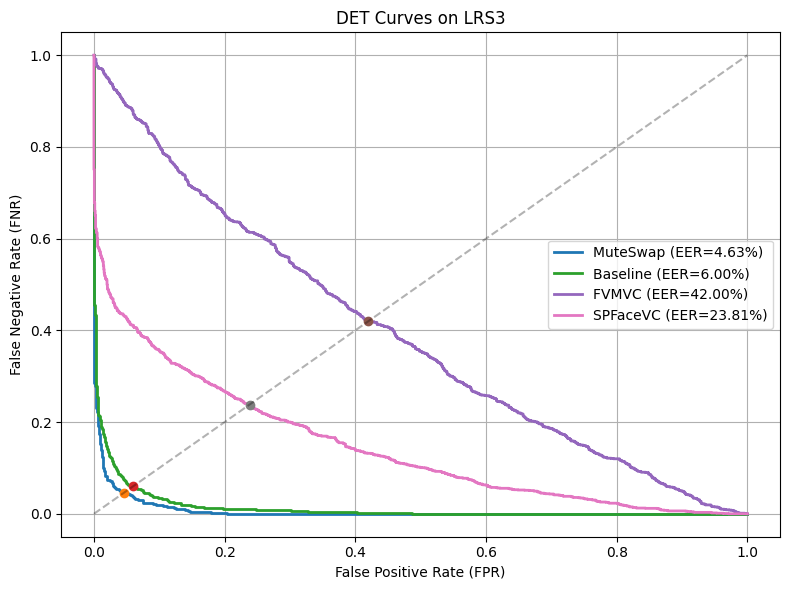} 
  \caption{DET Curve on LRS3 test pairs.}
  \label{fig:det_lrs3}
\end{figure}

Figure~\ref{fig:det_lrs3} presents the Detection Error Tradeoff (DET) curves for different voice conversion methods on LRS3 test pairs. Each curve reflects the tradeoff between false positive rate (FPR) and false negative rate (FNR) in speaker verification, with lower curves indicating better identity preservation.
MuteSwap achieves the best performance with the lowest Equal Error Rate (EER) of 4.63\%, outperforming the Baseline (6.00\%) and significantly surpassing prior FVC methods such as SP-FaceVC~\cite{SPFaceVC_AAAI23} (23.81\%) and FVMVC~\cite{FVMVC} (42.00\%). This demonstrates the effectiveness of MuteSwap in speaker verification despite operating without audio input.

The sharp contrast is especially evident in the steep slopes of FVMVC and SP-FaceVC curves, which show poor tradeoffs across all thresholds, implying weak alignment between the converted speech and target speaker identity. In contrast, MuteSwap maintains a low FNR even at low FPR, reflecting both precision and recall in identity conversion.
These results align with our quantitative evaluation and further confirm that MuteSwap enables accurate, perceptually consistent speaker identity synthesis from silent video, offering a viable alternative to audio-driven conversion methods.

\subsubsection{Identity Interpolation}

\begin{figure*}[ht]
  \centering
  \includegraphics[width=\linewidth]{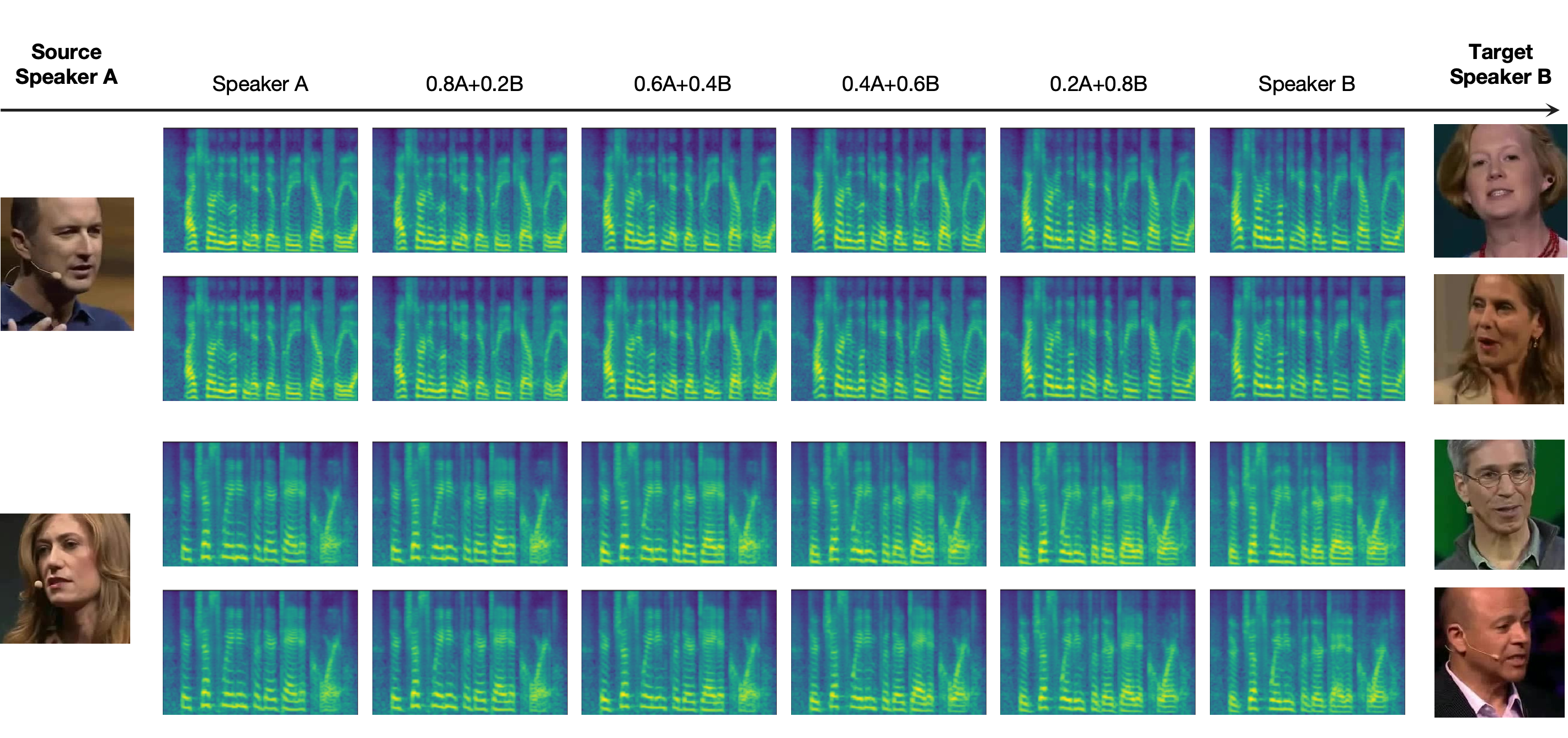} 
  \caption{Identity Embedding interpolation results. The mixing ratio for target speaker facial embedding increases from left to right. }
  \label{fig:sup_id_interp}
\end{figure*}

In this section, we exhibit more samples of identity embedding interpolation for MuteSwap. We can see in Figure~\ref{fig:sup_id_interp} that given different source and target speaker pairs, the conversion is done in a gradual manner as we increment the linear mixing ratio, which further validates our findings in the main paper.

\end{document}